\documentclass[11pt]{article}
\usepackage{moriond,epsfig}
\usepackage{subfigure}

\def\Journal#1#2#3#4{{#1} {\bf #2}, #3 (#4)}

\def\PLB{{\em Phys. Lett.}  B}
\def\PRL{\em Phys. Rev. Lett.}
\def\PRD{{\em Phys. Rev.} D}

\def\be{\begin{equation}}
\def\ee{\end{equation}}
\def\bea{\begin{eqnarray}}
\def\eea{\end{eqnarray}}

\newcommand{\ppbar}{p\bar{p}}
\newcommand{\ZZ}{ZZ}
\newcommand{\Dzero}{D\O\ }
\newcommand{\ZZllll}{\ZZ \rightarrow llll}

\begin{document}
\vspace*{4cm}
\title{Dibosons at the Tevatron}

\author{ Elliot Lipeles for the CDF and D\O\ Collaborations}

\address{CERN, CH-1211 Gen\`{e}ve 23, Switzerland}

\maketitle\abstracts{
  Recent developments in the study diboson production at the Tevatron
  are reviewed. These include indications at the 2.6$\sigma$ level for
  a radiation amplitude zero in the $W\gamma$ process at \Dzero and a 
  4.4$\sigma$ signal for $ZZ$ production in hadron collisions from CDF.
}

\section{Introduction}

The process of the simultaneous production of two electroweak bosons 
is one of the few tree-level processes that is sensitive to the couplings 
between gauge bosons. These couplings are a direct consequence of the
non-abelian group structure of the standard model (SM). 
At the Tevatron, a broad program of measuring cross-sections and kinematic 
distributions is aimed
testing whether these processes are consistent with the SM
predictions and searching for evidence of non-standard model contributions.
The $WW$ and $ZZ$ final states are also of interest as potential Higgs
search channels.

\begin{figure}
\begin{center}
\subfigure[\label{fig:t-channel}t-channel (a similar $u$-channel also contributes)]{\includegraphics[height=3cm]{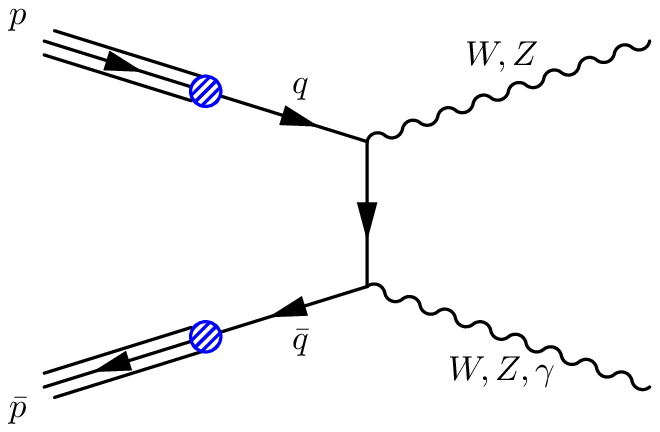}}
\hspace{0.25cm}
\subfigure[\label{fig:s-channel}s-channel]{\includegraphics[height=3cm]{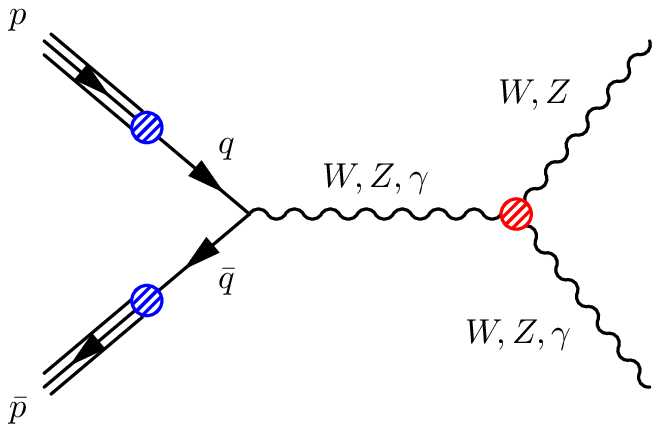}}
\hspace{0.25cm}
\subfigure[\label{fig:xsecplot} Cross-section summary as of March 2008]{\includegraphics[height=5cm]{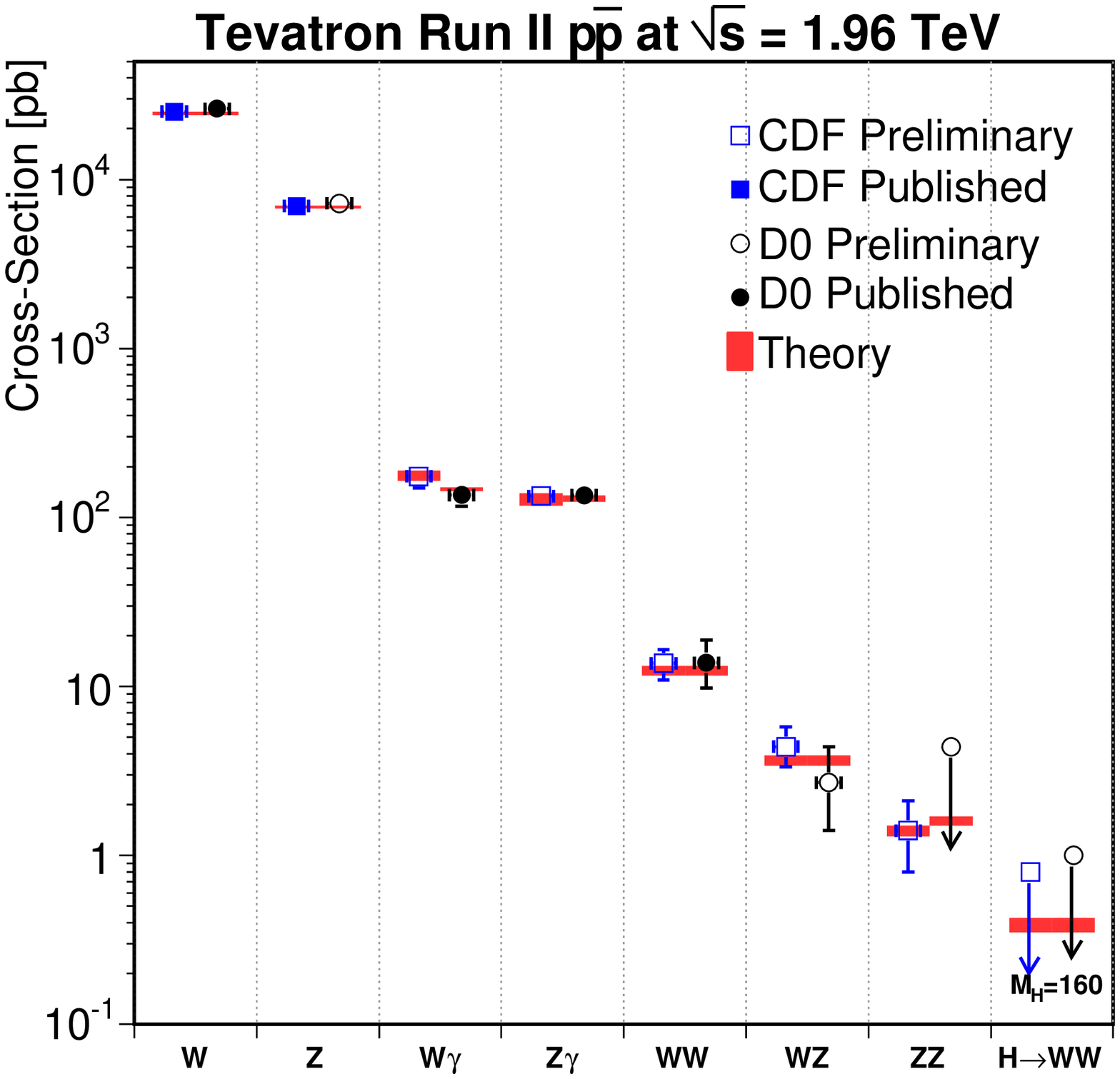}}
\end{center}
\caption{ Feynman diagrams contributing to diboson production at leading order and a summary of diboson cross-section measurements.
\label{fig:feyndiags}}
\end{figure}

The leading-order Feynman diagrams for diboson production are shown in 
Figure \ref{fig:feyndiags}. The $t$-channel shown in Figure \ref{fig:t-channel}
is effectively two copies of single boson production and involves only the
fermion to boson couplings. The $s$-channel shown in Figure \ref{fig:s-channel}
involves the triple gauge couplings. In the standard model, only the $WW\gamma$  
and $WWZ$ vertices are non-zero; the $Z\gamma\gamma$, $ZZ\gamma$, and $ZZZ$ vertices
do not exist.

The CDF and \Dzero have studied the $W\gamma$, $Z\gamma$, $WZ$, and
$ZZ$ final states in 1-2 fb$^{-1}$ of $\ppbar$ collisions at 1.96 TeV 
produced by the Tevatron. A summary of the predicted and observed cross-sections
is shown in Figure \ref{fig:xsecplot}. Shown for comparison are the single 
boson production cross-sections which present a significant experimental
challenge because they are three to four orders of magnitude 
larger than the diboson cross-sections.

\section{Radiation Amplitude Zero}

In the $W\gamma$ process, the two Feynman diagrams shown in Figure \ref{fig:feyndiags}
interfer destructively. This interference is complete when the angle of the $W^{\pm}$ relative
to the incoming quark in the $W\gamma$ rest-frame is $\pm\frac{1}{3}$ causing the 
leading-order differential cross-section
to completely disappear in what is known as the radiation amplitude zero (RAZ). Although
long predicted \cite{RAZ}, it is difficult to observe because the missing neutrino
information when $W$ is reconstructed in the $l\nu$ final state. There is however an
approximate zero in the quantity $Q_l \times (\eta_\gamma - \eta_l)$ where $Q_l$ is the lepton
charge, and $\eta_\gamma$ and $\eta_l$ are the pseudo-rapidities of the photon and lepton
respectively \cite{BaurRAZ}. The observation of the zero would be a demonstration of 
the presence of the $WW\gamma$ vertex contribution to the $W\gamma$ process.

\begin{figure}
\begin{center}
\subfigure[\label{fig:RAZa} Data compared to standard model]{\hspace{0.5cm}\includegraphics[height=4cm]{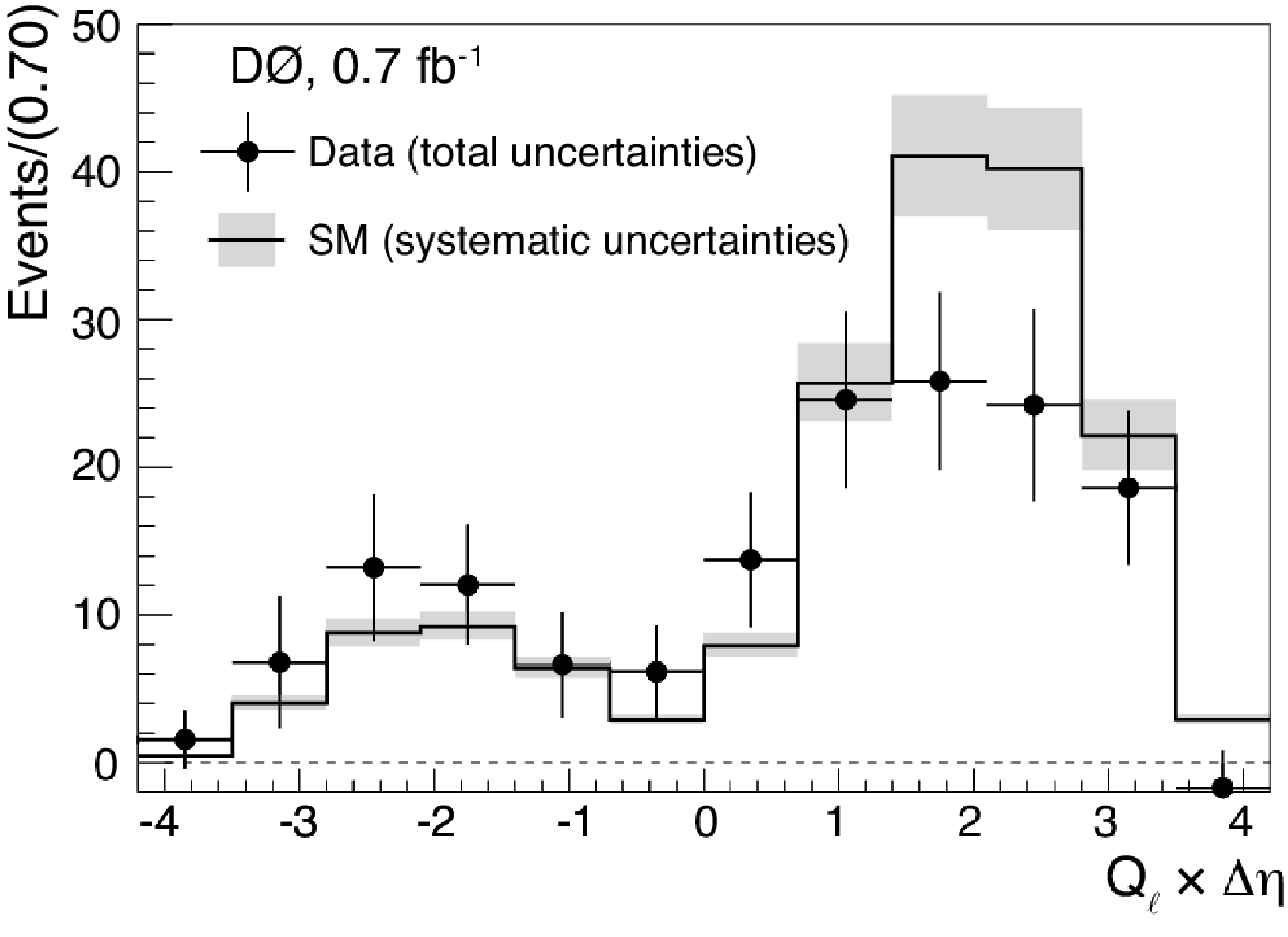}\hspace{0.5cm}}
\subfigure[\label{fig:RAZb} Models for dip significance calculation]{\hspace{0.5cm}\includegraphics[height=4cm]{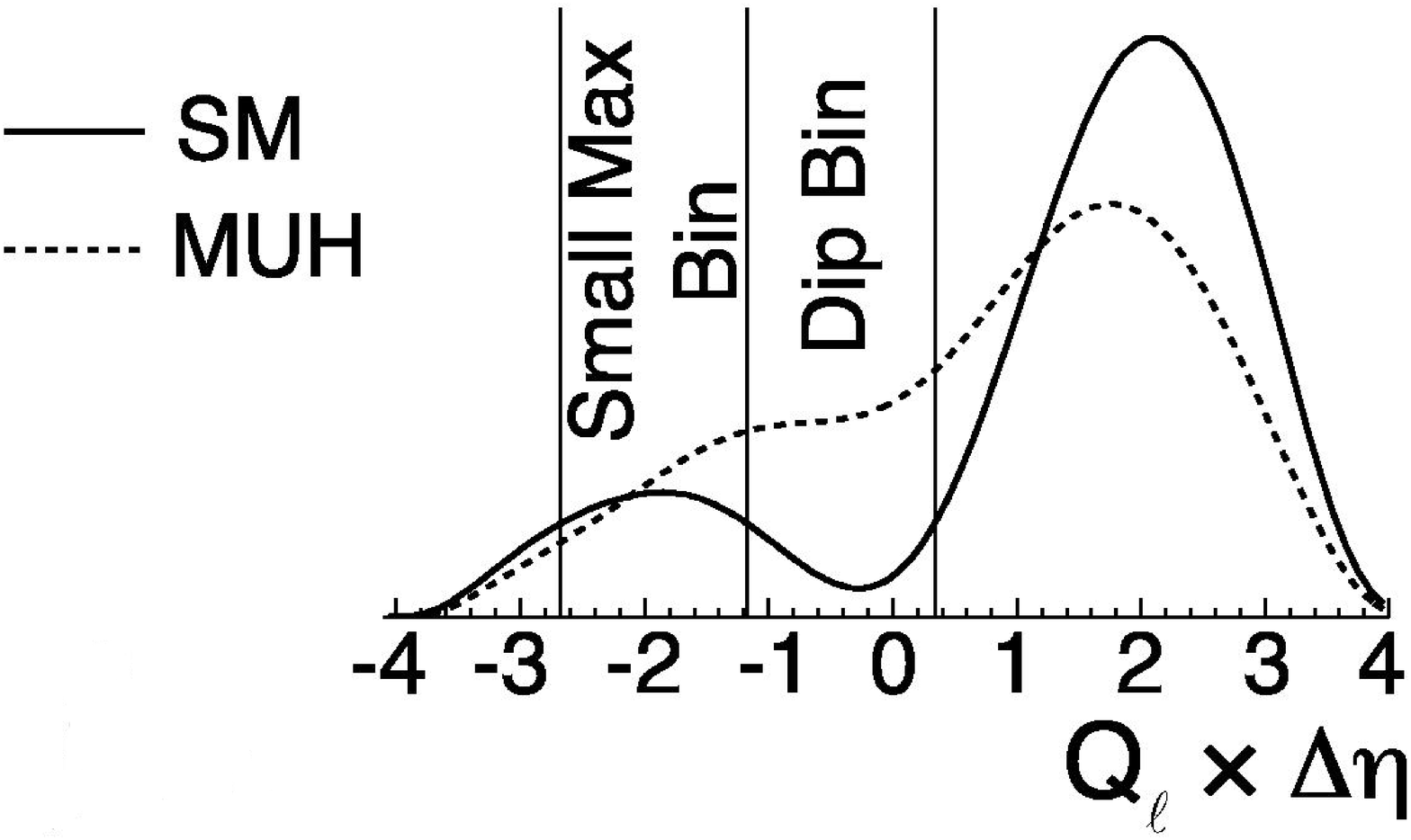}\hspace{0.5cm}}
\end{center}
\caption{The $Q_l \times (\eta_\gamma - \eta_l)$ distribution for \Dzero $W\gamma$ events.
\label{fig:RAZ}}
\end{figure}

Using 0.7 fb$^{-1}$ of $\ppbar$ collisions, \Dzero has reconstructed a sample of $W\gamma\rightarrow l\nu\gamma$
events where $l$ is either $e$ or $\mu$ \cite{DzeroRAZ}. The $Q_l \times (\eta_\gamma - \eta_l)$ distribution of 
these events, after subtracting the estimated background, it shown in Figure \ref{fig:RAZa}.  In order
to quantify the significance of the dip, a minimal unimodal hypothesis (MUH) is constructed by
choosing a set of $WW\gamma$ couplings such that there is no dip in the distribution (shown in Figure \ref{fig:RAZb}). 
They then find that it is ruled out at the 2.6$\sigma$ level.

\section{Triple Gauge Couplings}

\begin{figure}
\begin{center}
\subfigure[\label{fig:TGC_Wg} \Dzero $W\gamma$ photon $E_T$]{\includegraphics[height=4.45cm]{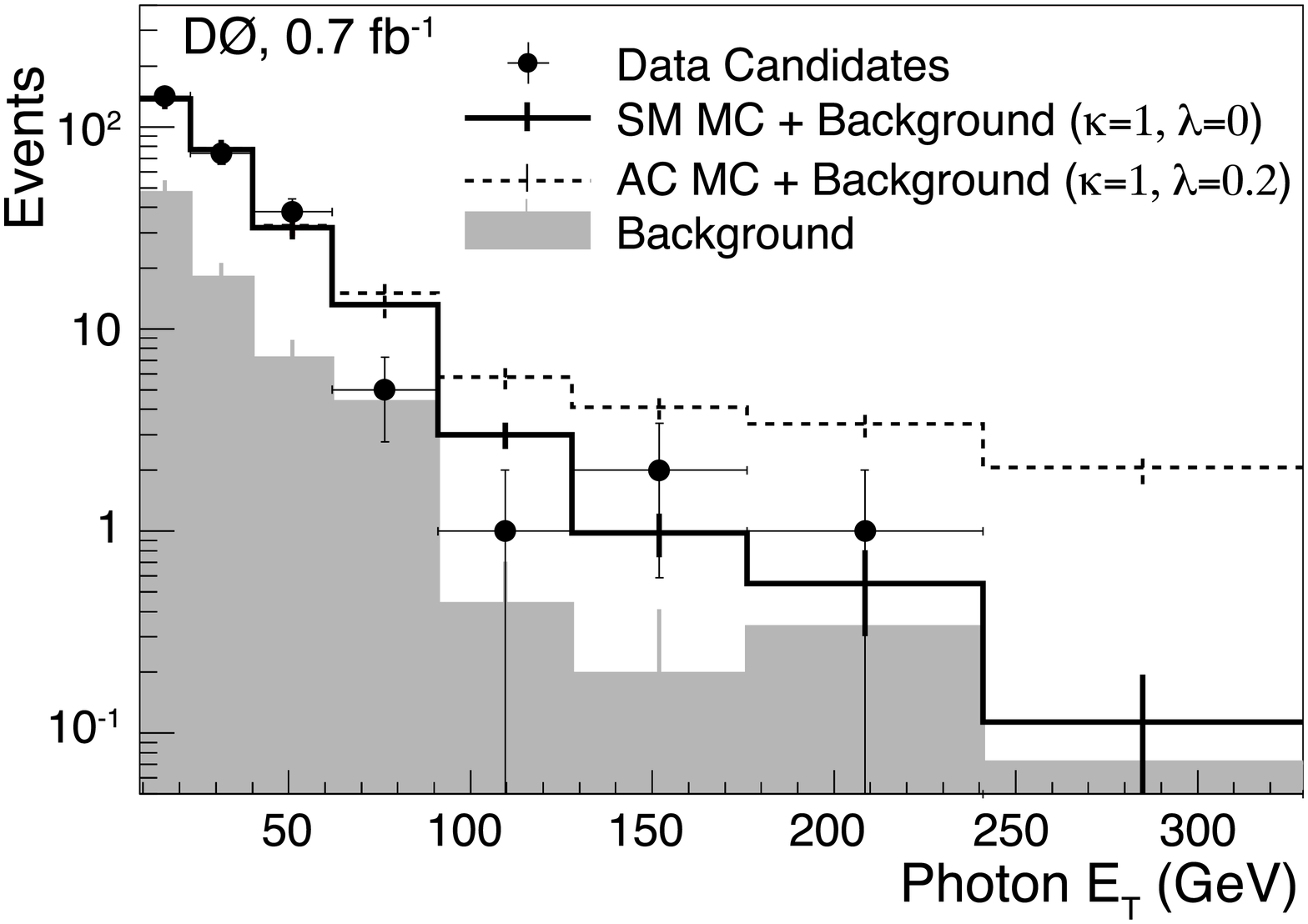}}
\subfigure[\label{fig:TGC_Zg} \Dzero $Z\gamma$ photon $E_T$]{\includegraphics[height=4.65cm]{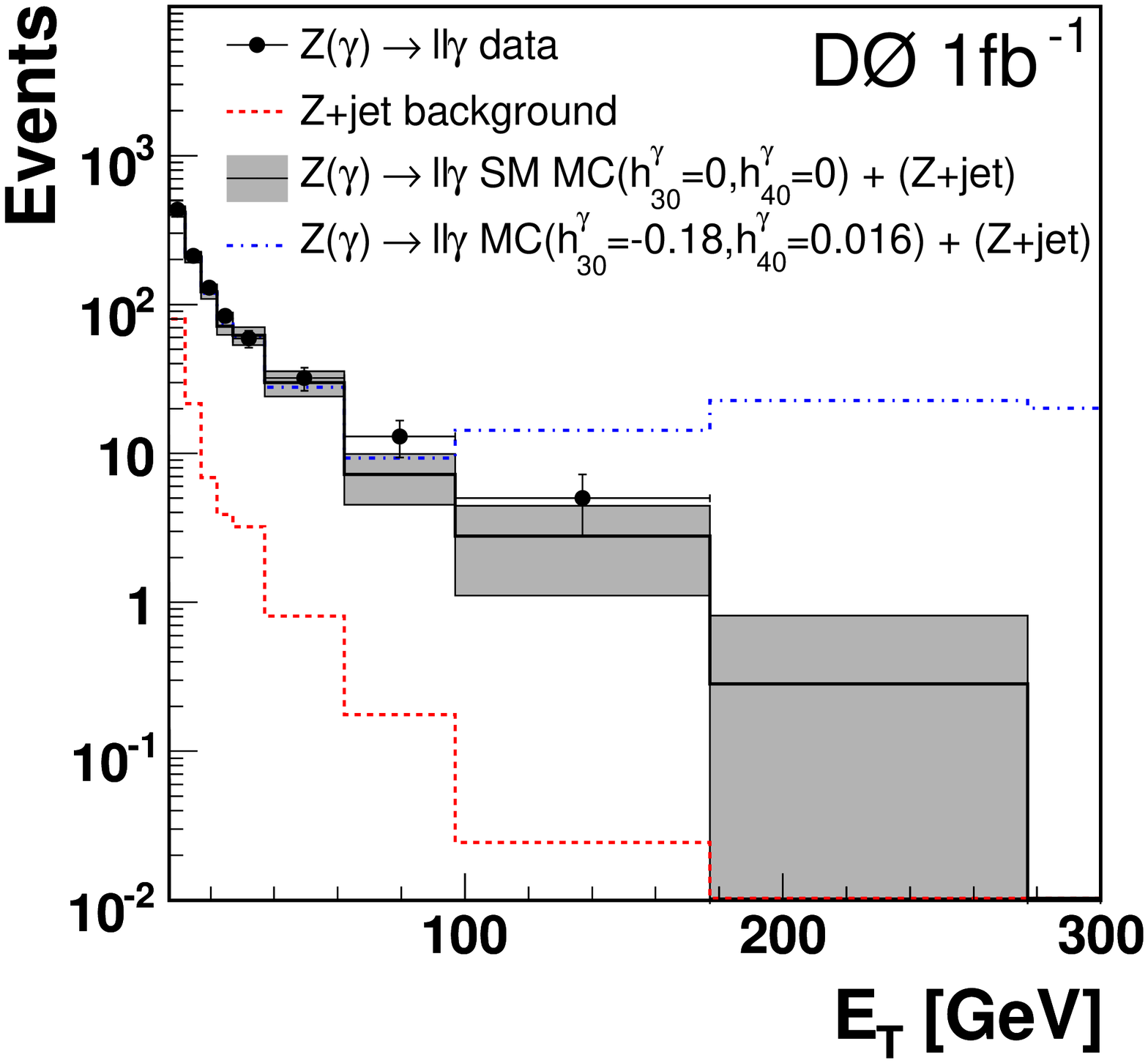}}
\subfigure[\label{fig:TGC_WZ} CDF $WZ$ $Z$-boson $p_T$]{\includegraphics[height=4.75cm]{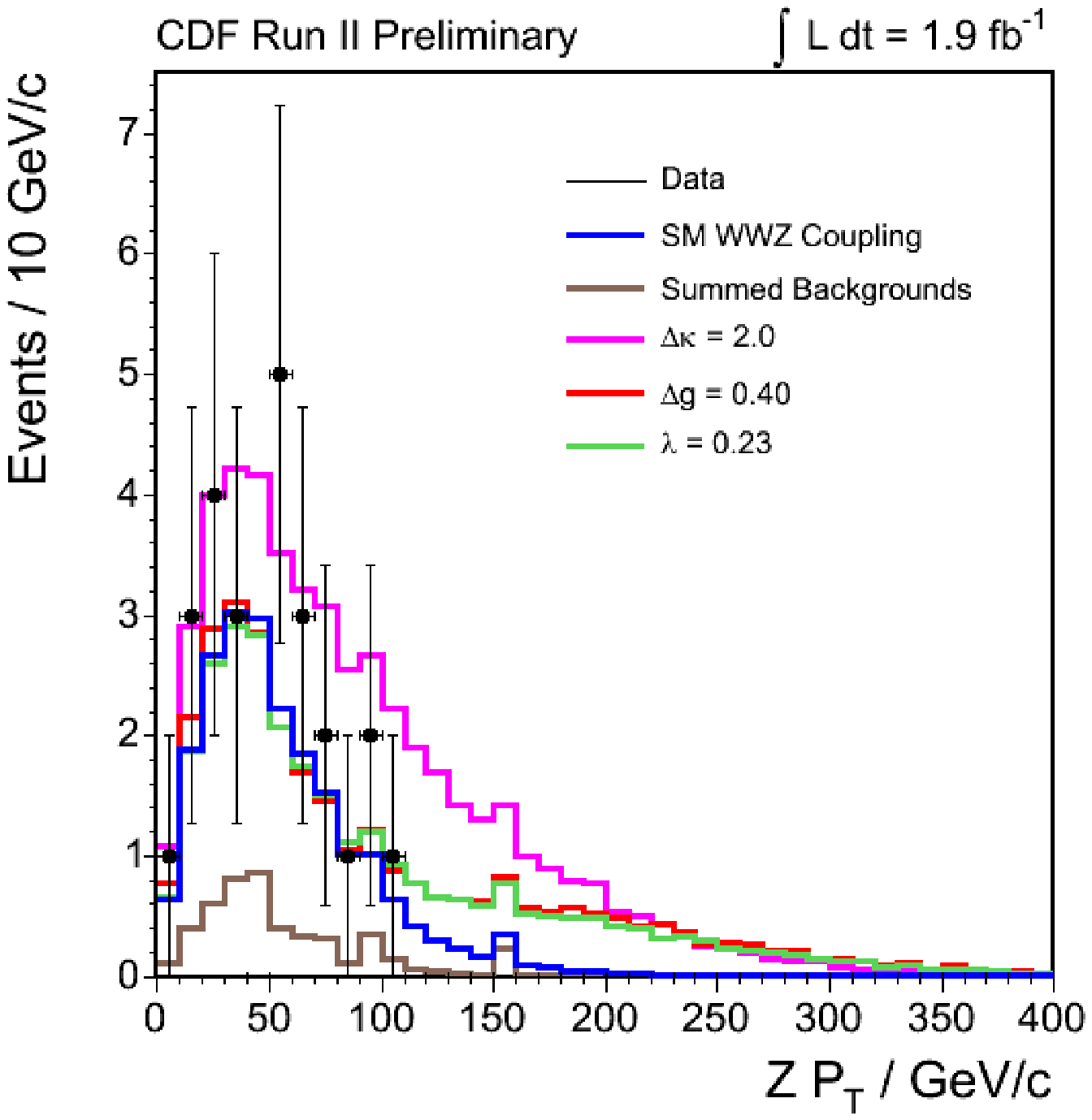}}
\end{center}
\caption{Distributions used to set limit on anomalous triple gauge couplings
\label{fig:TGC}}
\end{figure}

Deviations of the boson to boson couplings from the SM are referred to as 
anomalous triple gauge couplings (aTGCs) and are parameterized
by adding terms to the SM Lagrangian; for example
for the $WW\gamma$ vertex:
\begin{equation}
{\cal L}_{aTGC}/i g_{WW\gamma}
= \Delta\kappa_\gamma           W^*_{\mu}W_\nu F^{\mu\nu}                       
+ \frac{\lambda_\gamma}{M^2_W}  W^*_{\rho\mu}W^\mu_\nu F^{\nu\rho}
\end{equation}
where the form-factors $\lambda_\gamma$ and $\Delta\kappa_\gamma$ are zero in the SM. 
In addition to differences in the integrated cross-sections, anomalous TGCs typically 
give rise to significant enhancements at large diboson invariant mass $\hat{s}$.
In fact aTGCs can cause unitarity violations at large $\hat{s}$,
so the form-factors must be constructed so as to turn off as $\hat{s}$ gets large; e.g.
$\lambda_\gamma(\hat{s}) = \lambda_\gamma/(1+(\hat{s}/\Lambda)^2)^2)$ where $\Lambda$
is typically 1.5 to 2.0 TeV. This also means that these form-factors are intrinsically
energy dependent and Tevatron limits should be considered as complimentary to the LEP
limits which are at $\hat{s} \approx 2 M_{W}$. CDF and \Dzero
have recently updated limits on aTGCs using the $\gamma$ $E_T$  distributions
for $W\gamma$ (D\O \cite{DzeroRAZ})  and $Z\gamma$ (CDF \cite{CDFweb} and D\O\ \cite{DzeroZgam}), 
the $Z$ boson $p_T$ for $WZ$ (CDF \cite{CDFweb} and D\O\ \cite{DzeroWZ}),
and the cross-section alone for $ZZ$ (D\O\ \cite{DzeroZZ}) . Sample distributions 
used in setting aTGC limits are shown in Figure \ref{fig:TGC}.

\section{Evidence of $ZZ$ Production}

The $ZZ$ final state is the only SM diboson state not yet conclusively observed in 
hadron collisions (not including those involving the Higgs) and is unique in providing access to the $ZZZ$
coupling. \Dzero has searched in the four charged-lepton $llll$ channel \cite{DzeroZZ}
finding 1 candidate event in 1.0 fb$^{-1}$ of data
with an expected signal yield of  1.71$\pm$0.15 events and background of 0.13$\pm$0.03 events. Based of this 
search an upper limit of $\sigma(ZZ) < 4.4$ pb is set to be compared to an NLO prediction of 1.6 pb.

\begin{figure}
\begin{center}
\subfigure[Four lepton invariant mass distribution]{
 \hspace{1.0cm} \includegraphics[height=5cm]{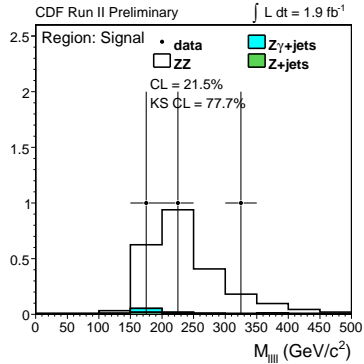}\hspace{1.0cm}}
\hspace{0.25cm}
\subfigure[Matrix-element discriminator for $ll\nu\nu$]{
  \hspace{1.0cm}\includegraphics[height=5cm]{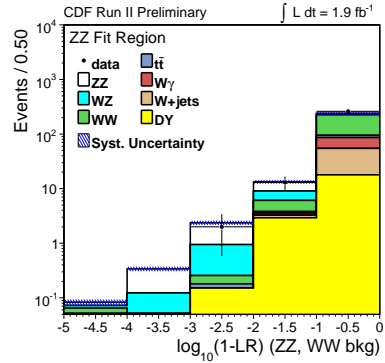}\hspace{1.0cm}}
\end{center}
\caption{Distributions from the CDF $ZZ$ search
\label{fig:ZZ}}
\end{figure}

CDF finds a 4.4$\sigma$ signal for $ZZ$ production \cite{CDFZZ} using 1.9 fb$^{-1}$ of data by combining
the $llll$ (4.2$\sigma$) and $ll\nu\nu$ (1.2$\sigma$) channels.
The $llll$ channel is subdivided into two categories based on whether the candidate contains
an electron that occurs outside the acceptance of the tracking system and therefore has a significantly
large background rate. Three $llll$ candidate events are found 
with the predicted signal and background yields shown in Table \ref{tab:CDF_llll}. In the $ll\nu\nu$ channel,
a matrix-element (ME) based probability calculation is used to separate the $ZZ$ signal from the much larger 
$WW$ background. The likelihood ratio from the ME calculation is shown in Figure \ref{fig:ZZ} along
with the four lepton invariant mass distribution from the $llll$ channel.
The measured cross-section $\sigma(p\overline{p}\rightarrow \ZZ) =
1.4^{+0.7}_{-0.6}$~(stat.+syst.)~pb is consistent with the standard
model expectation.
\begin{table}[t]
\caption{\label{tab:CDF_llll} Expected and observed number of $\ZZllll$ candidate events.
  The first uncertainty is statistical and the second one is systematic.
  }
\label{tbl:ZZllll_results}
\begin{center}
\begin{tabular}{lcc}
               & Candidates without a                             &     Candidates with a              \\
Category       & trackless electron                             &     trackless electron           \\  
               
\hline         
   $ZZ$             &   1.990 $\pm$ 0.013 $\pm$ 0.210           &     0.278 $\pm$ 0.005 $\pm$ 0.029      \\
   $Z$+jets       \parbox{0.0cm}{\vspace{0.55cm}}   &   ${0.014}^{+0.010}_{-0.007} \pm 0.003$   &  ${0.082}^{+0.089}_{-0.060} \pm 0.016$ \\
\hline
   Total          \parbox{0.0cm}{\vspace{0.65cm}}  &   $2.004^{+0.016}_{-0.015} \pm 0.210 $    &  $0.360^{+0.089}_{-0.060} \pm 0.033$   \\
\hline
{\bf Observed}      &             {\bf  2 }                           &      {\bf  1 }     \\
\end{tabular}
\end{center}
\end{table}

\section{Summary}

The increased luminosity at the Tevatron has allowed for substantial progress in diboson
physics and marks entry into a new sensitivity regime where electroweak bosons are now being 
pair produced in significant numbers. Recent accomplishments include 2.6$\sigma$ signal
for the RAZ in $W\gamma$, a 4.4$\sigma$ signal for $ZZ$ production, and limits on
anomalous couplings that continue to improve.




\end{document}